\documentclass[a4paper]{jpconf}
\usepackage{graphicx}
\usepackage{dcolumn}% Align table columns on decimal point
\usepackage{bm}% bold math
\usepackage{booktabs,graphicx,mathrsfs,verbatim}
\usepackage[colorlinks=true,citecolor=blue,filecolor=blue,linkcolor=blue,urlcolor=blue,pdftex]{hyperref}
\usepackage[usenames,dvipsnames]{color}
\usepackage{ulem}
\usepackage{color}

\newcommand{\1}[1]{\, \mathrm{#1}} 
\newcommand{\n}[1]{\mathrm{#1}} % normal (roman) text in math mode

\newcommand{\earth}{\oplus}

\begin{document}

\title{Results from the XENON100 Dark Matter Search Experiment}
\author{Laura Baudis} 
\address{Physics Institute, University of Z\"urich, CH-8057 Z\"urich, Switzerland}

\ead{laura.baudis@physik.uzh.ch (for the XENON collaboration)}

\begin{abstract}

{\small XENON100} is a liquid xenon time projection chamber built to search for rare collisions of hypothetical,  weakly interacting massive particles {\small (WIMPs)}, which are candidates for the dark matter in our universe, with  xenon atoms. Operated in a low-background shield at the Gran Sasso Underground Laboratory in Italy, {\small XENON100} has reached the unprecedented background level of $<$0.15\,events/(day$\cdot$keV) in the energy range below 100\,keV in 30\,kg of target mass, before electronic/nuclear recoil discrimination.  It found no evidence for {\small WIMPs} during  a dark matter run lasting for 100.9 live days in 2010,  excluding with 90\% confidence scalar {\small WIMP}-nucleon cross sections above  
7$\times10^{-45}\1{cm^2}$ at a {\small WIMP} mass of 50$\1{GeV/c^2}$.  A new run started in March 2011, and more than 210 live days of 
{\small WIMP}-search data were acquired. Results are expected to be released in spring 2012. The construction of the ton-scale {\small XENON1T}  detector in Hall B of the Gran Sasso Laboratory will start in late 2012.

\end{abstract}

\section{Introduction}

The {\small XENON100} \cite{Aprile:2009yh,Aprile:2011dd} experiment was built to search for interactions of massive, cold dark matter particles in liquid xenon.
The motivation for this search comes from our current understanding of the universe.
Cosmological observations ranging from the measured abundance of primordial elements to the precise  
mapping of anisotropies in the cosmic microwave background, to the study of the distribution of matter on galactic, extragalactic and the largest observed scales, to observations of high-redshift supernovae, have led to a so-called standard model of cosmology. 
In this  model, our universe is spatially flat and composed of $\sim$4\% atoms, $\sim$23\% dark matter and $\sim$73\% dark energy \cite{Komatsu:2010fb}.  
Understanding the nature of dark matter  poses a significant challenge to astroparticle physics, for its solution 
may involve new particles with masses and cross sections characteristic of the electroweak scale.
Such weakly interactive massive particles {\small (WIMPs)}, which would have been in thermal equilibrium with quarks and leptons in the hot 
early universe, and decoupled when they were non-relativistic, represent a generic class of dark matter candidates~\cite{Lee:1977ua}. 
Concrete examples are the lightest superpartner in supersymmetry with R-parity conservation \cite{Jungman:1995df}, and the lightest Kaluza-Klein particle, for instance  the first excitation of  the hypercharge gauge boson, in theories with universal extra dimensions  \cite{Hooper:2007qk}. Perhaps the most intriguing aspect of the {\small WIMP} hypothesis is the fact that it is testable by experiment. {\small WIMPs} with masses around the TeV scale are within reach of high-energy colliders and of direct and indirect dark matter detection experiments \cite{Bertone:2004pz}.

\section{Direct detection of WIMPs}

Dark matter particles might be detected via their elastic collisions with atomic nuclei in earthbound, low-background detectors \cite{Goodman:1984dc}. 
The differential rate for  elastic scattering can be expressed as  \cite{Lewin:1995rx}:

\begin{equation}
\frac {dR}{dE_R}=N_{T}
\frac{\rho_{h}}{m_{W}}
                    \int_{v_{\rm min}}^{v_{\rm max}} \,d \vec{v}\,f(\vec v)\,v
                     \,\frac{d\sigma}{d E_R}\,  
\label{eq1}
\end{equation}

\noindent
where $N_T$ is the number of target nuclei, $\rho_h$ is the local dark matter density in the galactic halo, $m_W$ is the {\small WIMP} mass,
$\vec v$ and $f(\vec v)$ are the {\small WIMP} velocity and velocity distribution function  in the Earth frame and ${d\sigma}/{d E_R}$ is the {\small WIMP}-nucleus differential cross section. The nuclear recoil energy is $E_R={{m_{\rm r}^2}}v^2(1-\cos \theta)/{m_N}$, where $\theta$ is the  scattering angle in the  center-of-mass frame, $m_N$ is the nuclear mass and $m_{\rm r}$ is the reduced mass. The minimum velocity is defined as  $v_{\rm min} = (m_N E_{th}/2m_{\rm r}^2)^{\frac{1}{2}}$, where $E_{th}$ is the energy threshold of the detector, and $v_{\rm max}$ is the escape velocity in the Earth frame.  The simplest galactic model assumes a Maxwell-Boltzmann distribution for the WIMP velocity in the galactic rest frame, with a velocity dispersion of $\sigma \approx$ 270\,km\,s$^{-1}$ and an escape velocity of v$_{esc} \approx$ 544\,km\,s$^{-1}$.  For  direct detection experiments, the mass density  and velocity distribution at a radius around 8\,kpc are most relevant. High-resolution, dark-matter-only 
simulations of Milky Way-like halos find that the dark matter mass distribution at the solar position is smooth,  with substructures   being far away from the Sun. The local velocity distribution of dark matter particles is likewise found to be smooth, and  close to Maxwellian \cite{Vogelsberger:2008qb}. Recent simulation of hierarchical structure formation including the effect of baryons revealed that a thick dark matter disk forms in galaxies, along with the dark matter
halo \cite{lake08,Read:2009iv}. The dark disk has a density of $\rho_{d}/\rho_{h}$ = 0.25$-$1.5 where the standard halo density is $\rho_h$=0.3\,GeV\,cm$^{-3}$ and the kinematics are predicted to follow  the Milky Way's stellar thick disk. At the solar neighborhood, this yields a rotation lag of v$_{lag}$=40$-$50\,km/s with respect to the local circular velocity, and a dispersion of  $\sigma \simeq$40$-$60\,km/s.  These velocities are significantly lower than in the standard halo model  and have implications  for the expected rates in direct \cite{Bruch:2008rx}  and indirect \cite{Bruch:2009rp} dark matter detection experiments.

The differential cross section for elastic scattering has two separate components: an effective scalar coupling between the {\small WIMP} and the mass of the nucleus  and  an effective coupling between the spin of the {\small WIMP} and the total angular momentum of the nucleus. In general, the coherent part dominates the interaction (depending however on the characteristics and composition of the dark matter particle) for target masses with A$\geq$30 \cite{Jungman:1995df}. The total cross section  is  the sum of both contributions    
$\frac{d\sigma}{d E_R} \propto {\sigma_{SI}^0} F_{SI}^2(E_R) + {\sigma_{SD}^0} F_{SD}^2(E_R)$, where  $\sigma_{SI,SD}^0$ are the spin-independent (SI) and  spin-dependent (SD)  cross sections in the limit of zero momentum transfer and $F_{SI,SD}^2(E_R)$ denote the nuclear form factors, expressed as a function of the recoil energy.
These become significant at large WIMP and nucleus masses, leading to a suppression of the differential scattering rate at higher recoil energies. 

A dark matter particle with a mass in the GeV$-$TeV range  has a momentum of a few tens to a few hundred MeV and an energy below 100\,keV is transferred to a nucleus in a terrestrial detector. Expected event rates range from  one to less than 10$^{-3}$  events per kg detector material and year. To observe a {\small WIMP}-induced spectrum, a low energy threshold, an ultra-low background noise and a large target mass are essential. In a given detector, the kinetic energy carried by the scattered nucleus is transformed into a measurable signal, such as ionization, scintillation light or phonons. The simultaneous detection of two observables  
yields a powerful discrimination against background events,  which are mostly interactions with electrons, as opposed to {\small WIMPs} and neutrons, which  scatter  off nuclei.  Given the size and rate of a potential dark matter signal, an absolute low background is equally relevant. It implies high material selection of detector components and   specific active and passive shields against the natural radioactivity of the environment  and against cosmic rays and secondary particles produced in their interactions.  Highly granular detectors and/or good timing and position resolution will distinguish localized energy depositions from multiple scatters within the active detector volume and in addition allow to  intrinsically  measure the neutron background. Finally, the position resolution leads to the identification of events clustered at the detector surfaces or elsewhere,  which are quite unlikely to be induced by {\small WIMP} interactions.

\section{Liquid xenon dark matter detectors}

Liquid xenon has ideal properties as a dark matter target and is used to build massive, 
homogeneous and position-sensitive {\small WIMP} detectors. It has a high scintillation ($\lambda$ = 178\,nm) 
and ionization yield because of its low ionization potential of 12.13\,eV  \cite{elena-book}.
Scintillation is produced by the formation and radiative decay of so-called excimers, Xe$_2^*$, which are bound ion-atom states. Due to the deexcitation of the singlet and triplet states of the excited dimers, the scintillation decay times have a fast ($\sim$4.2\,ns) and a slow ($\sim$22\,ns)  component.  Although the intensity ratio of the singlet to triplet state depends on the deposited energy density, the effect is difficult to exploit in practice, in particular at low recoil energies, because of the similar involved time scales\footnote{This effect is widely explored in liquid argon detectors, where the decay times of the singlet and triplet states are 7\,ns and 1.6\,$\mu$s, respectively.}. 
If an electric field around 1\,kV/cm is applied, ionization electrons can also
be detected, either directly or through the secondary process of proportional scintillation.
The interaction of a {\small WIMP} results in a low-energy nuclear recoil,  which loses its energy generating charge carriers and scintillation photons. 
Both signals are suppressed when compared to an electronic recoil of the same initial energy, but by different amounts, allowing 
to use their ratio to distinguish between electronic and nuclear recoils. The  ionization and scintillation yields, defined as the number of produced electron-ion pairs and photons per unit of absorbed energy,  depend on the drift field and on the energy of the recoil and must be known precisely down to low energies. 
For electronic recoils, measurements and predictions for the light yield exist down to 5\,keV \cite{szydagis11}, while for nuclear recoils the light and charge yield were measured directly down to 3\,keV \cite{plante11,Aprile:2008rc} and 4\,keV \cite{manzur10}, respectively.
Natural xenon does not contain long-lived radioactive isotopes apart from the double beta emitter $^{136}$Xe, which  
has a half-life  of 2.1$\times$10$^{21}$\,yr \cite{exo_2011}. The trace content of $^{85}$Kr,  a beta emitter with  
an endpoint of 687\,keV and a half-life of 10.76 years, has to be separated to a level of 10$^{-12}$ mol/mol ($^{nat}$Kr/Xe) for a ton-scale experiment. 
It is achieved by distillation \cite{Abe:2008py}, which exploits the different boiling points of krypton (120\,K) and xenon (165\,K) at 1 atmosphere,  
 or by adsorption-based chromatography \cite{elena-book}.  These tiny krypton concentrations in xenon gas can be observed with the mass spectrometry technique \cite{dobi11}, which in the future might also be used to monitor the krypton levels in a dark matter detector. 
The high mass of the xenon nucleus is favorable for scalar interactions
and the presence of two isotopes with unpaired neutrons ($^{129}$Xe: spin-1/2, 26.4\% and $^{131}$Xe: spin-3/2, 21.2\%) 
ensures sensitivity to axial {\small WIMP}-nuclei couplings. The high density (3\,g/cm$^3$) and high atomic number (Z=54, A=131.3)  allow to build self-shielding, compact dark matter detectors. 

\section{The XENON program}

{\small XENON} is a phased approach to dark matter direct detection with time projection chambers {\small (TPCs)}  at the 10\,kg, 100\,kg and 1000\,kg fiducial target mass scale.  
An interaction within the active volume of the detector will create ionization electrons and prompt scintillation photons, and  both signals are detected. The electrons drift in the pure liquid under an external electric field, are  accelerated by a stronger field and extracted into the vapour phase above the liquid, where they generate proportional scintillation, or electroluminiscence.  Two arrays of photomultiplier tubes, one in the liquid and one in the gas, detect the prompt scintillation (S1) and the delayed proportional scintillation signal (S2). The array immersed in the liquid collects the majority of the prompt signal, which is totally reflected at the liquid-gas interface. The ratio of the two signals is different for nuclear recoils  created by {\small WIMP} or neutron interactions, and electronic recoils produced by $\beta$ and $\gamma$-rays, providing the basis for background discrimination. 
Since electron diffusion in the ultra-pure liquid xenon is small, the proportional scintillation photons carry the $x-y$ information of the interaction site. With the $z-$information from the drift time measurement, the {\small TPC} yields  a three-dimensional event localization, enabling to reject the majority of the background via fiducial volume cuts. 
To test the concept and verify the achievable energy threshold, background rejection and dark matter sensitivity, a detector with a fiducial mass on the order of 10\,kg {\small (XENON10)} 
\cite{Angle:2007uj,xenon10_instrument}, was developed and operated at the Gran Sasso National Laboratory {\small (LNGS)}. {\small  XENON10} excluded previously unexplored  parameter space, setting  90\% C.L. upper limits of  $\rm 4.5 \times 10^{-44}\,cm^{2}$  and  $\rm 5 \times 10^{-39}\,cm^{2}$ on the {\small WIMP}-nucleon  spin-independent and spin-dependent cross-sections, respectively, at a {\small WIMP} mass of 30~GeV/$c^{2}$  \cite{Angle:2007uj,Angle:2008we}. It found no evidence for light ($\le$10\,GeV) dark matter particles with scattering cross sections above 10$^{-42}$\,cm$^{2}$ \cite{xenon10prlLDM}. 
The next step in this program, the XENON100 detector \cite{Aprile:2011dd}, is taking data at {\small LNGS} since 2008.  I will briefly describe the experiment and its recent results in the following sections. {\small XENON1T}, with a total liquid xenon mass of  2.4\,tons, is approved to be built in Hall B at {\small LNGS}, in a 9.6\,m diameter and 10\,m height water Cherenkov shield \cite{xenon1t_tdr}. Construction will start in late 2012. Finally, {\small DARWIN} \cite{darwin_2010,darwin_2011,darwin} is a design study for  an ``ultimate''   noble liquid dark matter experiment, aiming to probe cross sections down to 10$^{-48}$cm$^2$ and to measure {\small WIMP}-induced nuclear recoil spectra with high-statistics, should they be discovered by an existing or near-future experiment. The  goal is to determine or at least  constrain  {\small WIMP} properties, such as its mass, scattering cross section and possibly spin \cite{Pato:2010zk}. 

\section{The XENON100 detector}

{\small XENON100} is a 161\,kg double-phase xenon {\small TPC} operated at LNGS in an improved {\small XENON10} shield \cite{Aprile:2009yh,Aprile:2011dd}. A total of 242 low-radioactivity, UV-sensitive photomultiplier tubes {\small (PMTs)} detect the prompt  and proportional  light signals induced by particles interacting in the sensitive volume, containing 62\,kg of ultra-pure liquid xenon. The remaining 99\,kg act as an  active veto shield mostly against multiple Compton scatters. While the fiducial mass of {\small XENON100} has been increased by a factor of 10 with respect to {\small XENON10}, its background is lower by a factor of 100 \cite{aprile2011vb}. This remarkable reduction was achieved through screening and selection of ultra-low background materials for the detector and shield construction \cite{Aprile:2011ru}, by placing the cryogenic system along with its cryocooler and the high-voltage feedthroughs outside of the shield, by taking advantage of the self-shielding power of xenon and of the active xenon shield, by purifying the xenon for the radioactive $^{85}$Kr with a dedicated distillation column operated underground, and by adding $\sim$5\,cm of electrolytic copper and 20\,cm of water inside and outside of the existing shield, respectively.

\section{Detector design}

The design of the {\small XENON100} detector and present performance results are detailed in \cite{Aprile:2011dd}.
The time projection chamber  is close to cylindrical in shape,  with 30.6\,cm in diameter and 30.5\,cm in height. Its  walls are  made of 24 interlocking  polytetrafluorethylen (teflon) panels,  which work 
as insulators and are good reflectors for the xenon scintillation light ~\cite{ref::yamashita2004}. The electrical fields are created with thin metal 
meshes that were etched with a hexagonal pattern from stainless steel foils and spot-welded onto low-radioactivity, stainless steel rings. The cathode mesh, which is 75\,$\mu$m thick  and has a pitch of 5\,mm, 
is biased with 16\,kV, generating a drift field of 0.53\,kV/cm across the {\small TPC}. About 5\,mm below the top {\small PMT} array, the {\small TPC} is closed with a
stack of three  meshes:  a central anode, 125\,$\mu$m thick and  2.5\,mm pitch is placed between two grounded meshes with a spacing of 5\,mm. An electron extraction field of
$\sim$12\,kV/cm is obtained by applying $+$4.5\,kV to the anode.  The field is sufficiently high to obtain an electron extraction efficiency close to 100\%~\cite{ref::aprile2004}.  
Averaged over all angles of incidence, the optical transparency of the top mesh stack and of cathode plus an additional screening mesh below the cathode is 47.7\% and 83.4\%, respectively.
A homogeneous electric field across the drift region is created by 40~equidistant field shaping electrodes, connected through 700\,M$\Omega$ resistors. 

Two arrays of Hamamatsu R8520-06-AL 1-inch square {\small PMTs} with synthetic quartz windows and selected for low radioactivity in $^{238}$U/$^{232}$Th/$^{40}$K/$^{60}$Co/$^{137}$Cs ~\cite{Aprile:2011ru}, 
detect the {\small VUV} scintillation light:   98 tubes are located in the vapour  phase above the liquid target, arranged in concentric circles to optimize the spatial resolution of radial event-position reconstruction. While these tubes  mostly see the S2 light signal, the energy threshold of the detector is determined by the much smaller S1 signal. Because  of the 1.7 refractive index of LXe \cite{ref::solovov2004} and the resulting internal reflection at the  liquid-gas interface, about 80\% of the S1~signal is seen by a second {\small PMT} array immersed  in the liquid below the cathode. These  80 tubes have higher quantum efficiencies compared to the ones on top, and are arranged in a  closed-packed geometry, yielding optimal area coverage for efficient S1 light collection.

\begin{figure}[t]
  \centering
  \includegraphics*[width=0.38\textwidth]{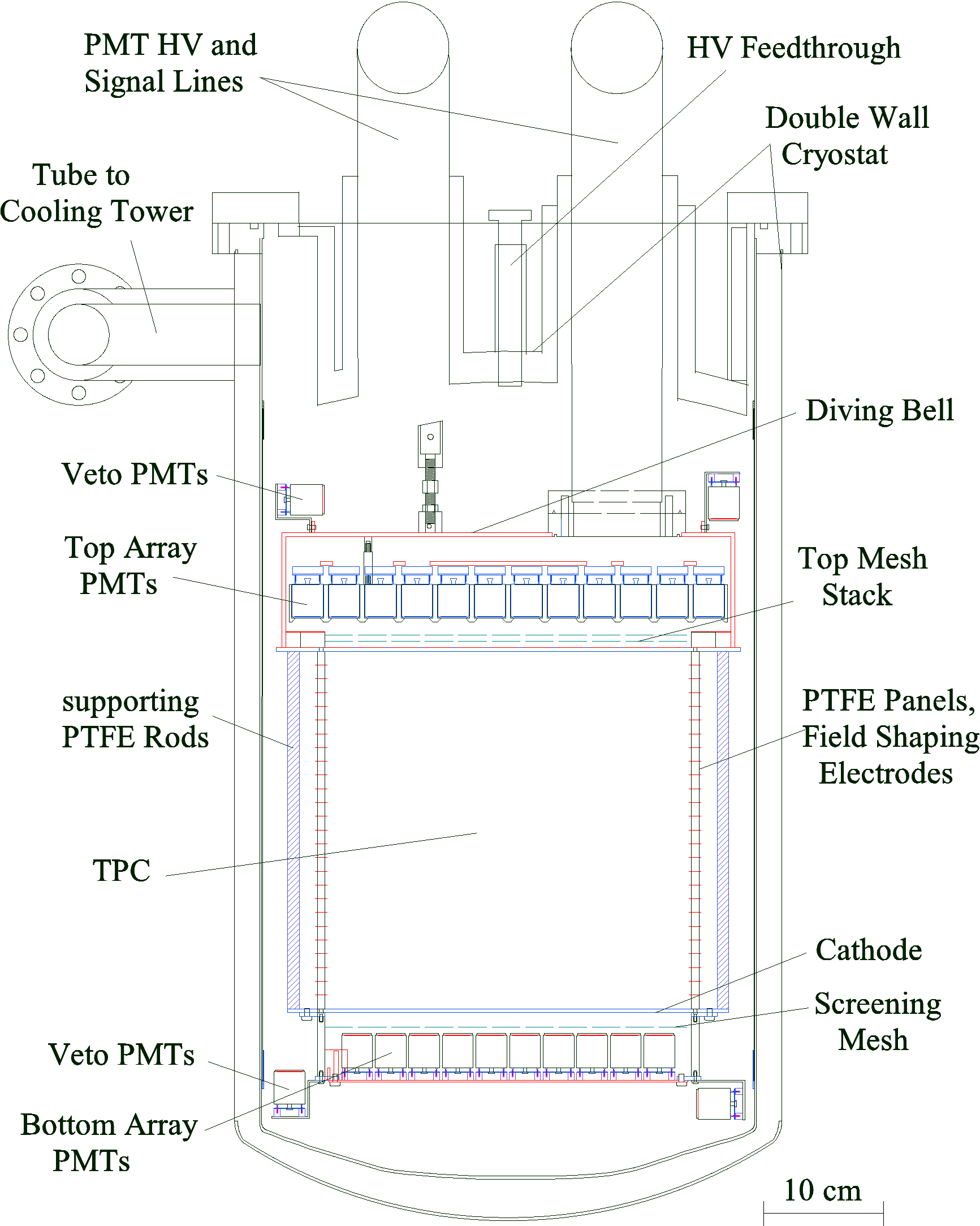}
  \includegraphics*[width=0.42\textwidth]{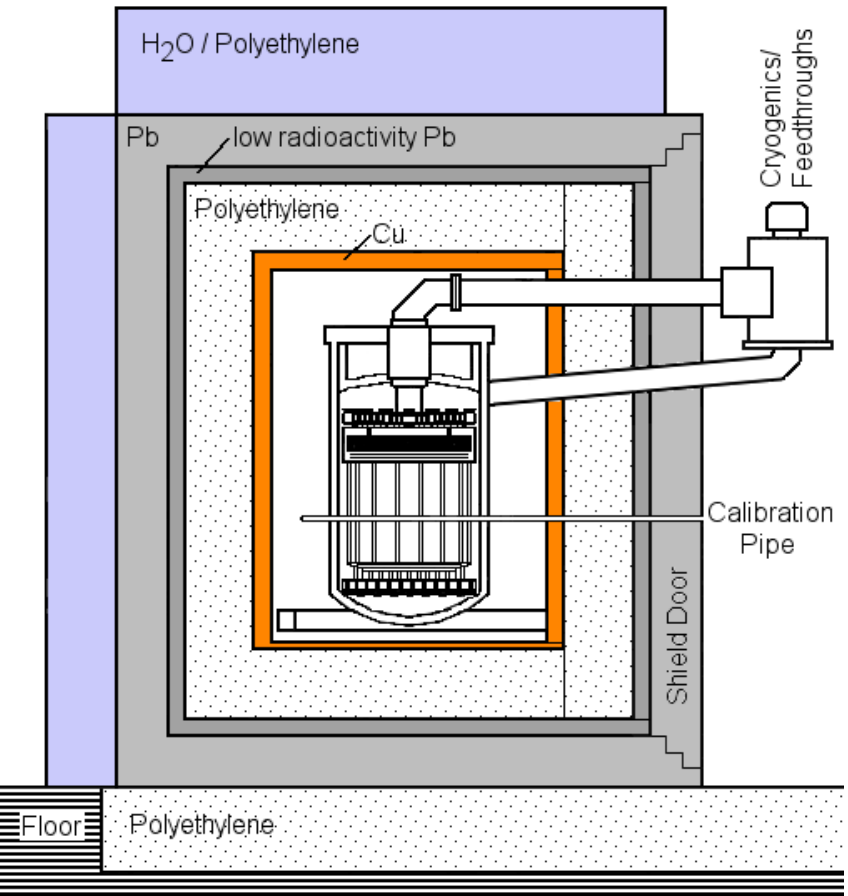}
  \caption{{\small (Left): Drawing of the XENON100 detector. The time projection chamber contains
62\,kg of liquid xenon and is surrounded by an active liquid xenon veto of 99\,kg. 
(Right): Schematic view of  XENON100  in its passive shield made of
copper, polyethylene, lead and water  \cite{Aprile:2011dd}.} }
\label{fig::xe100schematics}
\end{figure}

The {\small TPC} is enclosed in a double-walled, low-radioactivity stainless steel cryostat \cite{Aprile:2011ru}.  The detector is cooled remotely and the connection to the outside of the shield is established via three stainless steel pipes, one to the cooling system, two to the {\small PMT} feedthroughs and pumping ports.  To bias the cathode and the anode, custom-made hermetic high-voltage feedthroughs, composed of a stainless steel core enclosed by a teflon insulation layer, are used.  A liquid xenon layer of  4\,cm thickness surrounds the {\small TPC}  and  is observed by 64~{\small PMTs}. The active veto is used to reject multiple-scatter events occurring within a time window of $\pm$20~ns, and 
is  most effective in reducing the background from Compton scatters \cite{aprile2011vb}.  It is optically separated from the {\small TPC} by the interlocking teflon panels, reducing the event rate in 
the dark matter target and the rate of accidental coincidences to a negligible level.
A drawing of the detector and a schematic view  in its low-background shield are shown in figure \ref{fig::xe100schematics}.

\section{Detector calibration}

To characterize the detector performance and its stability in time, calibration sources are regularly inserted in the {\small XENON100} shield through a copper tube surrounding the cryostat 
(visible in figure \ref{fig::xe100schematics}, right). The electronic recoil band in $\log_{10}(\textrm{S2/S1})$ versus energy space defines the region of background events from $\beta$- and $\gamma$-particles. It is  measured using the low-energy Compton tail from  $^{60}$Co and $^{232}$Th 
calibration sources.  The detector response to single-scatter nuclear recoils, the expected signature of a dark matter particle, 
  is measured with an AmBe $(\alpha,n)$-source.  Besides the definition 
of the nuclear recoil band in $\log_{10}(\textrm{S2/S1})$ and thus of the  included {\small WIMP}-search region, the calibration yields gamma lines from inelastic neutron collisions, as well as from the de-excitation of xenon or fluorine (in the teflon) isomers, activated by neutron capture:
40\,keV from $^{129}$Xe, 80\,keV from $^{131}$Xe, 110\,keV from $^{19}$F, 164\,keV from $^{131m}$Xe (T$_{1/2}$=11.8\,d), 197\,keV from $^{19}$F, and 236\,keV from $^{129m}$Xe (T$_{1/2}$=8.9\,d).

\begin{figure}[b!]
  \centering
  \includegraphics*[width=0.40\textwidth]{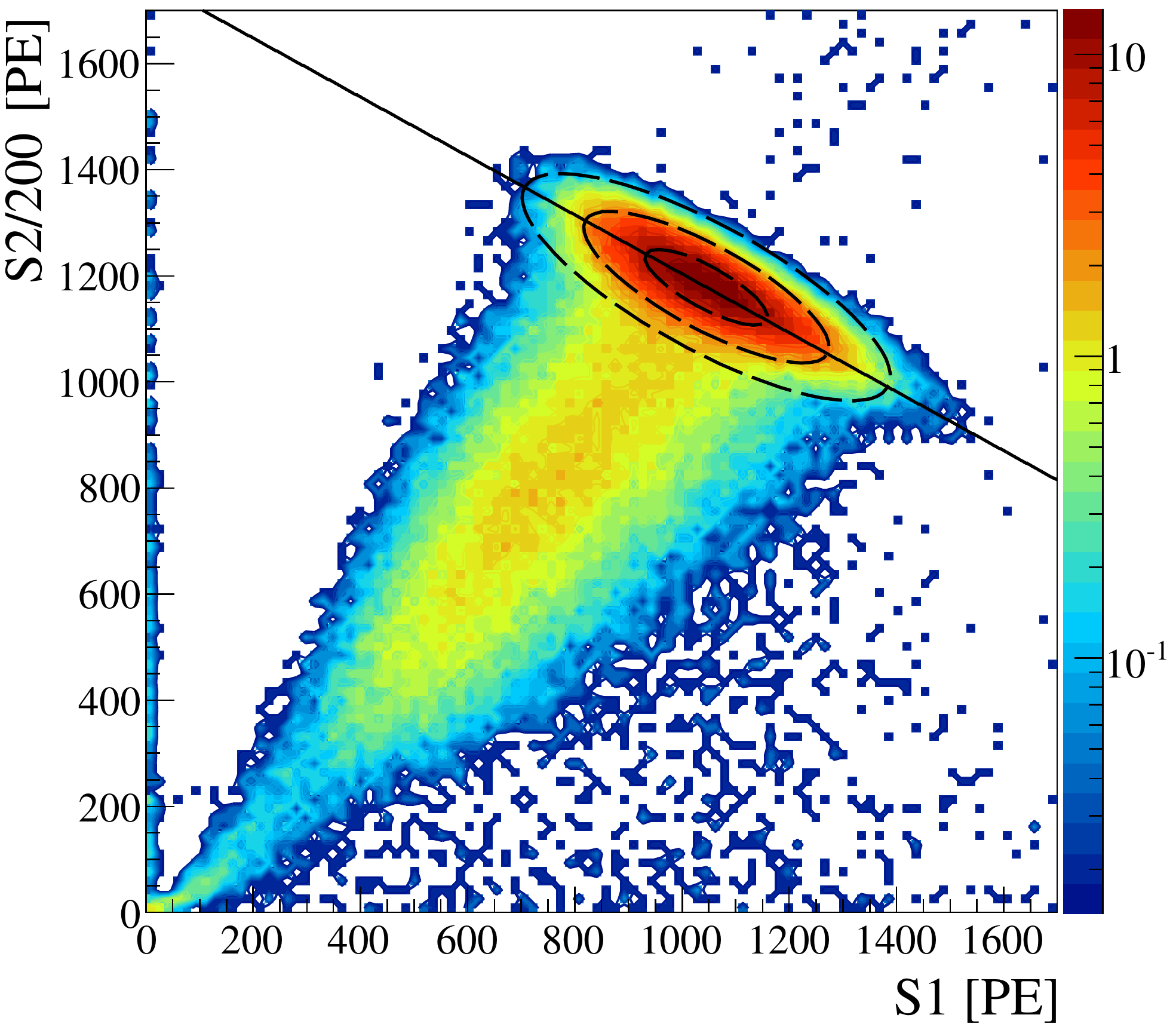}
   \includegraphics*[width=0.5\textwidth]{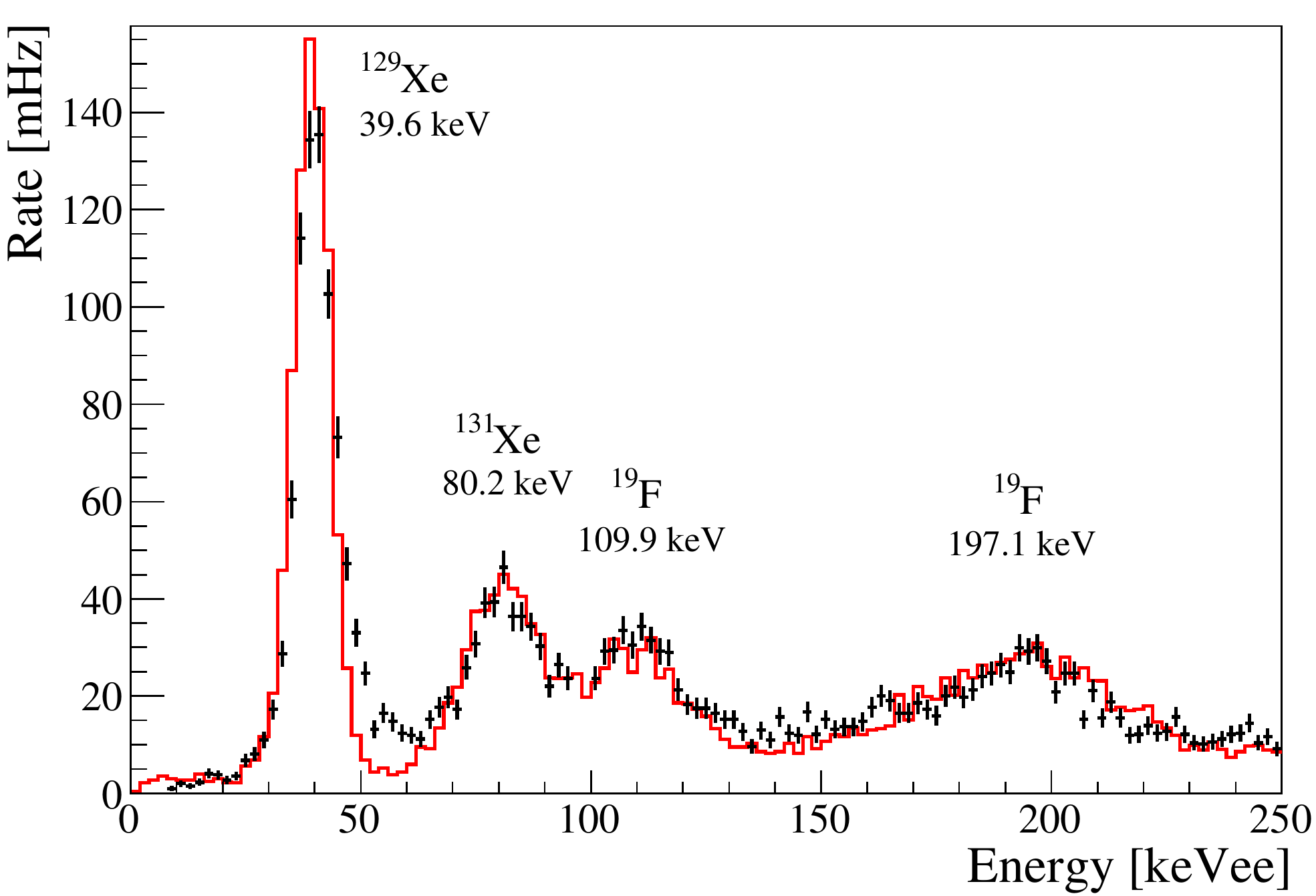}
  \caption{{\small (Left):  Data from  a $^{137}$Cs source in the S2-S1 parameter space.  The charge and light signals are anti-correlated, a projection along the anti-correlation ellipse 
leads to an improved energy resolution. (Right): AmBe calibration spectrum for inelastic neutron scatter in the combined energy scale, along with  Monte Carlo generated data (solid) \cite{Aprile:2011dd}.}}
\label{fig::anticorr}
\end{figure}

The ionization and scintillation signals are anti-correlated for interactions in LXe \cite{ref::ces_shutt,ref::ces_aprile,ref::ces_aprile2}. The fluctuations in the sum signal are lower than in each individual signal, leading to an improved energy resolution.  
 In {\small XENON100},  each calibration line  generates an ellipse in the S2-S1 plane, that can be described with a two-dimensional Gaussian  to determine the anti-correlation angle~$\theta$,  as shown in figure~\ref{fig::anticorr}, left.  This  angle has been determined for energies from 40\,keV to 1333\,keV: It is roughly constant for energies down to $\sim$100\,keV and decreases at lower energies. The angles at 40\,keV and 80\,keV are smaller, since the observed energy deposition is a combination from a nuclear recoil and a subsequent, prompt gamma emission. 
From the mean positions and angles, the so-called combined energy scale for electronic recoils is defined, and its linearity verified by comparing the resulting spectra to Monte Carlo data. An example is shown in figure~\ref{fig::anticorr}, right, where the measured electronic recoil spectrum from an AmBe calibration is compared to a Monte Carlo generated spectrum. This energy scale is currently used for background studies  \cite{aprile2011vb} alone,  the {\small WIMP} search data is analyzed with an S1-based nuclear recoil energy scale \cite{xenon100_prl107}.

\begin{figure}[tb]
  \includegraphics*[width=0.48\textwidth]{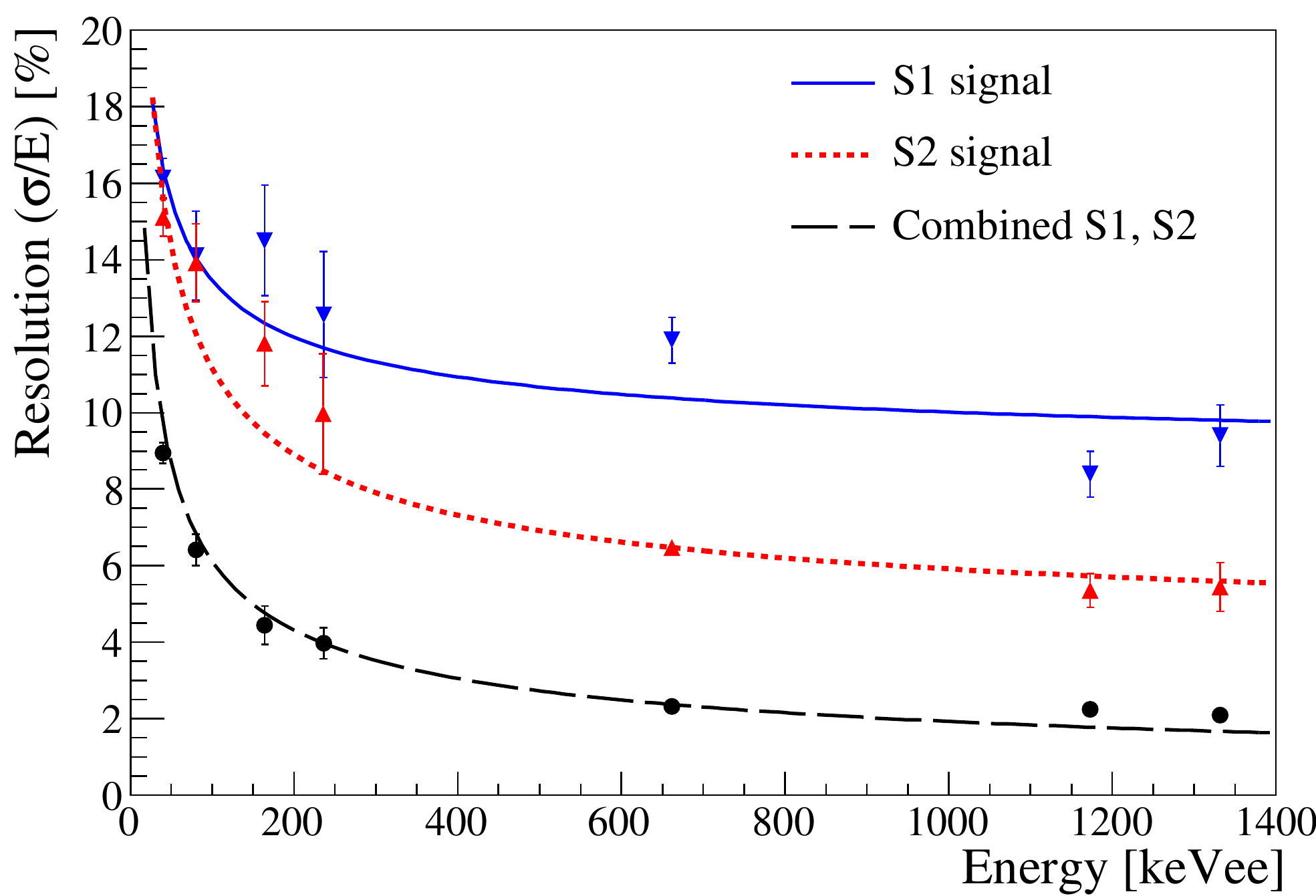}
  \includegraphics*[width=0.48\textwidth]{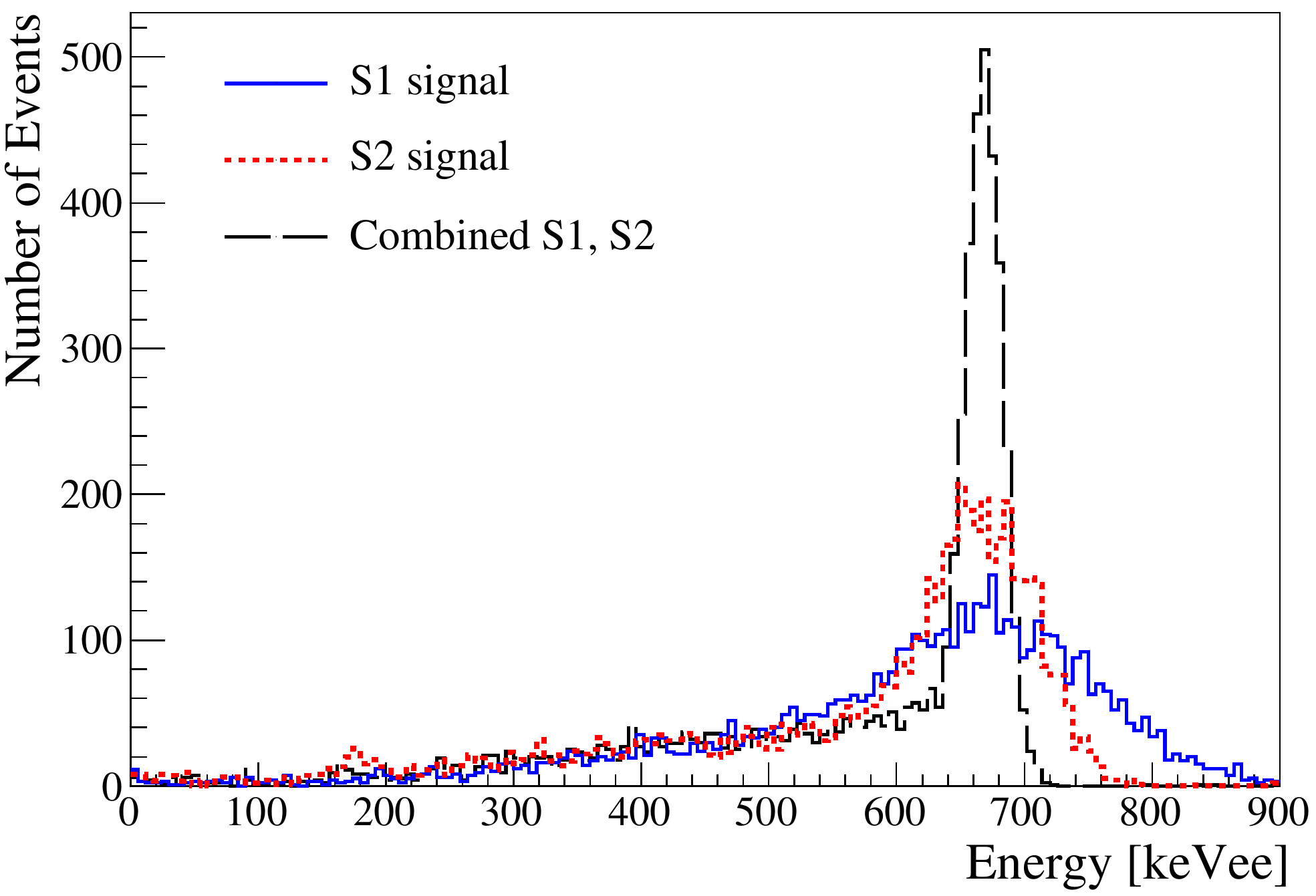}
  \caption{{\small (Left): Measured energy resolution using $\gamma$-lines between 40\,keV and 1333\,keV in S1, S2, and combined energy scale, along with fits to the $1/\sqrt{E}$ dependence. (Right): Spectrum of $^{137}$Cs at 662\,keV. The 1-$\sigma$ resolution is 12.5\%, 6.5\%  and  2.3\% for the S1,  S2, and the combined scale, respectively. Shown are single-scatters surviving an active veto cut which effectively reduces the Compton continuum \cite{Aprile:2011dd}. }}
 \label{fig::eres_lines}
\end{figure}

The energy resolution as a function of energy, as determined for three different scales, is shown figure~\ref{fig::eres_lines}, left. 
At 1\,MeV, the resolution is 12.2\%, 5.9\% and  1.9\% for the S1, the  S2, and the combined energy scale, respectively. 
Figure~\ref{fig::eres_lines}, right, illustrates the measured change in the $^{137}$Cs $\gamma$-peak using the three energy scales. 

\section{Backgrounds}
\label{sec:backgrounds}

The background predictions in {\small XENON100}  are based on Monte Carlo simulations of both electronic and nuclear recoil components, including the muon-induced neutron background \cite{alex_phd_thesis} under a rock overburden of $\sim$3600 m water equivalent. The simulations use as input a detailed geometry and the measured activities of all detector and shield construction materials \cite{aprile2011vb}. These materials  were selected for their low intrinsic radioactivity with a dedicated screening facility consisting of a 2.2\,kg high-purity germanium detector in an ultra-low background copper cryostat and a Cu/Pb shield, operated underground at LNGS \cite{Baudis:2011am}. While the nuclear recoil background is sub-dominant at the current sensitivity level (see section \ref{sec:results}), the electronic recoil part is dominated by interactions of $\gamma$-rays from decays of $^{238}$U, $^{232}$Th, $^{40}$K, and $^{60}$Co in detector materials, mostly in the {\small PMTs}, followed by the cryostat. The self-shielding and the active xenon veto reduce the rate significantly in the central part of the target: by a factor of  $\sim$20 and $\sim$40 for 40\,kg and 30\,kg of target mass,  compared to the total active xenon mass of 62\,kg.

During the commissioning run in fall 2009~\cite{xenon100_prl105}, the level of krypton in the liquid xenon has been measured with a delayed-coincidence technique using a decay channel where $^{85}$Kr $\beta$-decays to $^{85m}$Rb ($\tau$~=~1.46\,$\mu$s, E$_{max}$~=~173.4\,keV), which transitions to the ground state emitting a 514\,keV photon. The obtained concentration of $^{nat}$Kr/Xe was 143$_{-90}^{+130}\times$10$^{-12}$\,mol/mol,   assuming a $^{85}$Kr/$^{nat}$Kr abundance of 10$^{-11}$. The $^{222}$Rn level in the liquid has been determined using a $\beta-\alpha$ time-coincidence analysis, where events corresponding to the decays of $^{214}$Bi (T$_{1/2}$~=~19.7~min, E$_{max}$~=~3.27~MeV) and $^{214}$Po (T$_{1/2}$~=~164~$\mu$s, E$_{\alpha}$~=~7.69~MeV) are tagged. The derived upper limit on the $^{222}$Rn activity in liquid xenon was 21~$\mu$Bq/kg.

\begin{figure}[!t]
\includegraphics[height=0.33\linewidth]{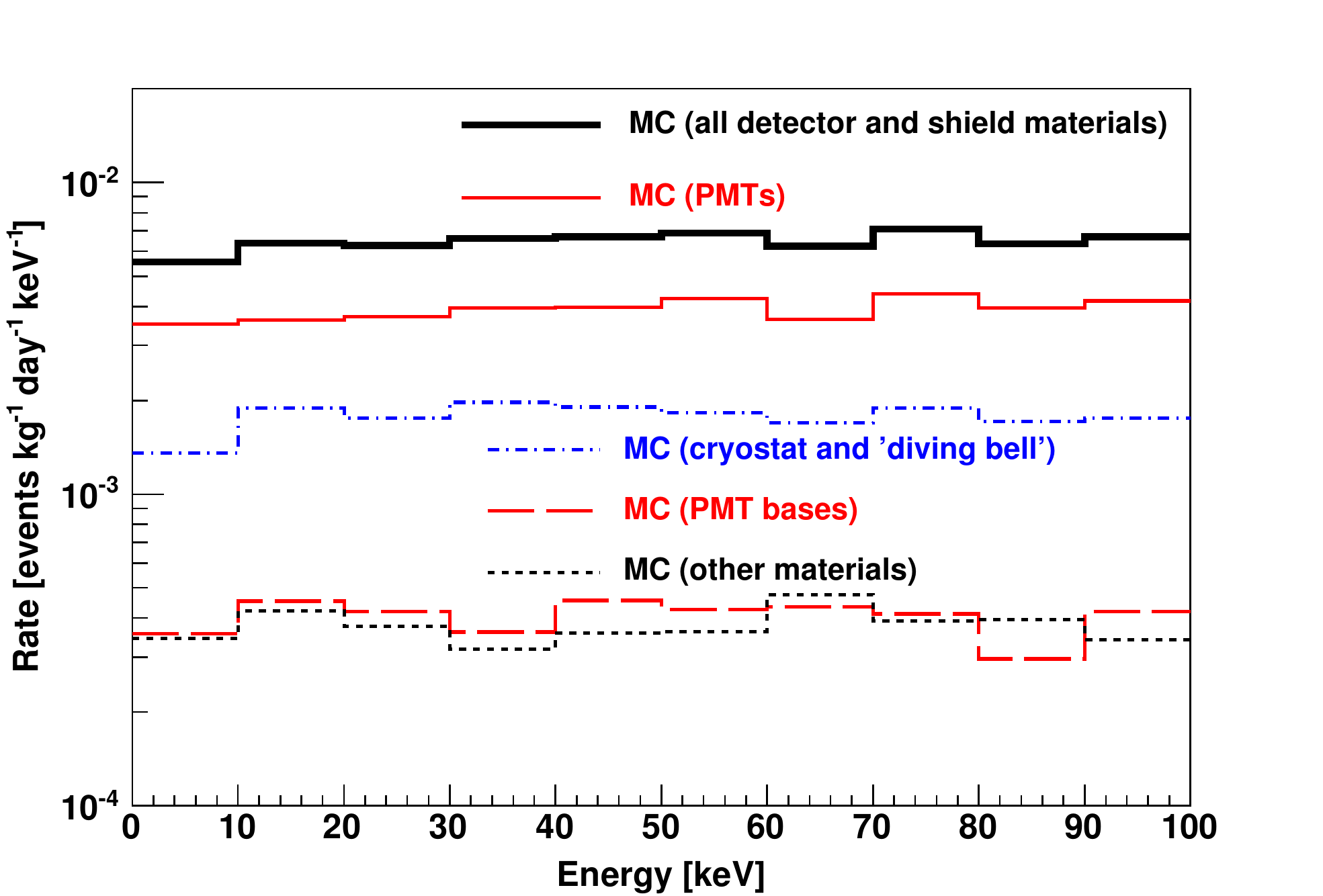}
\includegraphics[height=0.33\linewidth]{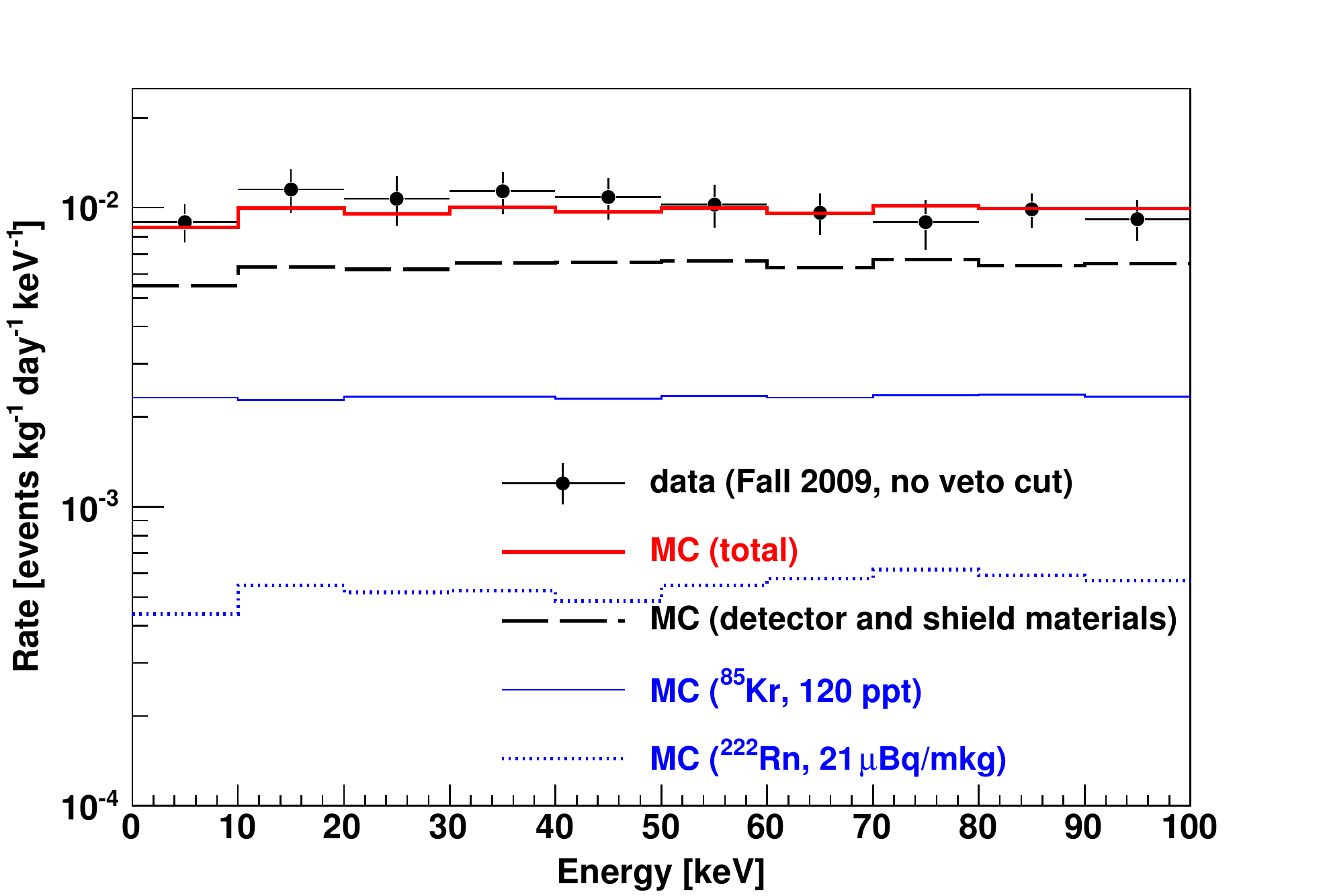}
\caption{{\small (Left): Predicted background from detector and shield materials (thick solid) in a 30\,kg fiducial target mass without veto cut,  along with  individual contributions from the {\small PMTs} (solid), the cryostat (dashed-dotted), the  {\small PMT} bases (long dashed), and all  remaining detector components (short dashed). (Right):  Zoom into the low-energy region of figure~\ref{figDataMC}: measured and Monte Carlo generated background spectra in a 30\,kg fiducial volume before an active veto cut \cite{aprile2011vb}.}}
\label{figDataMCzoom}
\end{figure}

\begin{figure*}[!ht]
\begin{center}
\includegraphics[width=0.95\linewidth]{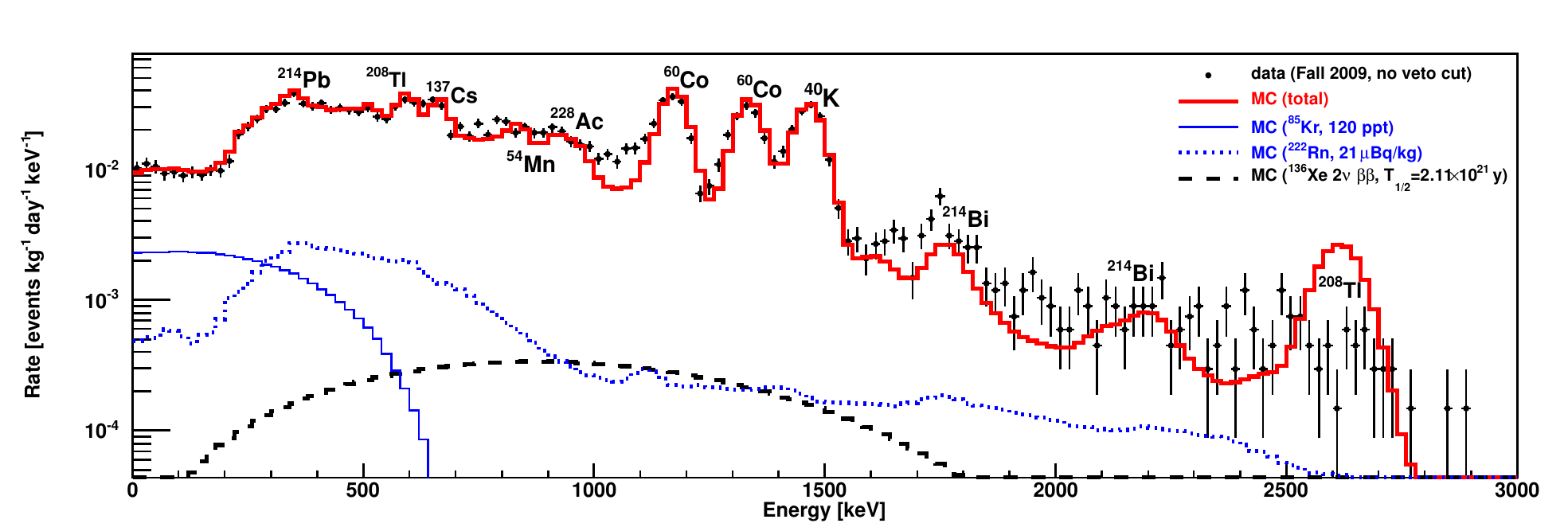}
\caption{{\small Measured electronic recoil background spectrum during the commissioning run \cite{xenon100_prl105}, along with the total Monte Carlo generated spectrum (solid) in a 30\,kg fiducial xenon volume before an active veto cut.  The energy spectra from $^{85}$Kr decays (thin solid), $^{222}$Rn decays (dotted) and the recently measured 2$\nu$ double beta decay spectrum of $^{136}$Xe (dashed) are also shown \cite{aprile2011vb}. }}
\label{figDataMC}
\end{center}
\end{figure*}

A comparison of the measured electronic recoil background spectrum and the Monte Carlo prediction for a central target region of 30\,kg, before an active veto cut, is shown in figure~\ref{figDataMC}. A zoom into the low-energy region is shown 
in figure~\ref{figDataMCzoom}, right.  Excellent agreement of the background model with the data is observed in the  energy region below 700~keV, and for the main $\gamma$-peaks. 
 In particular, simulated and measured background spectra agree well in the energy region of interest for the dark matter search. 
 The disagreement at higher energies is caused by  non-linear effects in the {\small PMT} response, which affects the performance of the position reconstruction algorithms, changing the event rate in the fiducial volumes and leading to a degradation of the position-dependent signal corrections \cite{aprile2011vb}.

 The total predicted rates of single-scatter electronic recoil events in the energy region of interest are 0.63 (0.24) events/(day$\cdot$keV) and 0.29 (0.14) events/(day$\cdot$keV) for 40\,kg and 30\,kg target mass, without (with) an active veto cut, respectively. In the 30\,kg fiducial target mass, $^{85}$Kr  decays make about 55\% of the veto-anticoincident background,  while the contribution from $^{222}$Rn in the liquid  is $<$7\%.  An electron recoil discrimination level of 99.5\%, based on the S2/S1-ratio,   reduces this background by an additional factor of 200.

\section{Science run and results}
\label{sec:results}

\begin{figure}[htbp]
%\centering
\includegraphics[width=0.48\columnwidth]{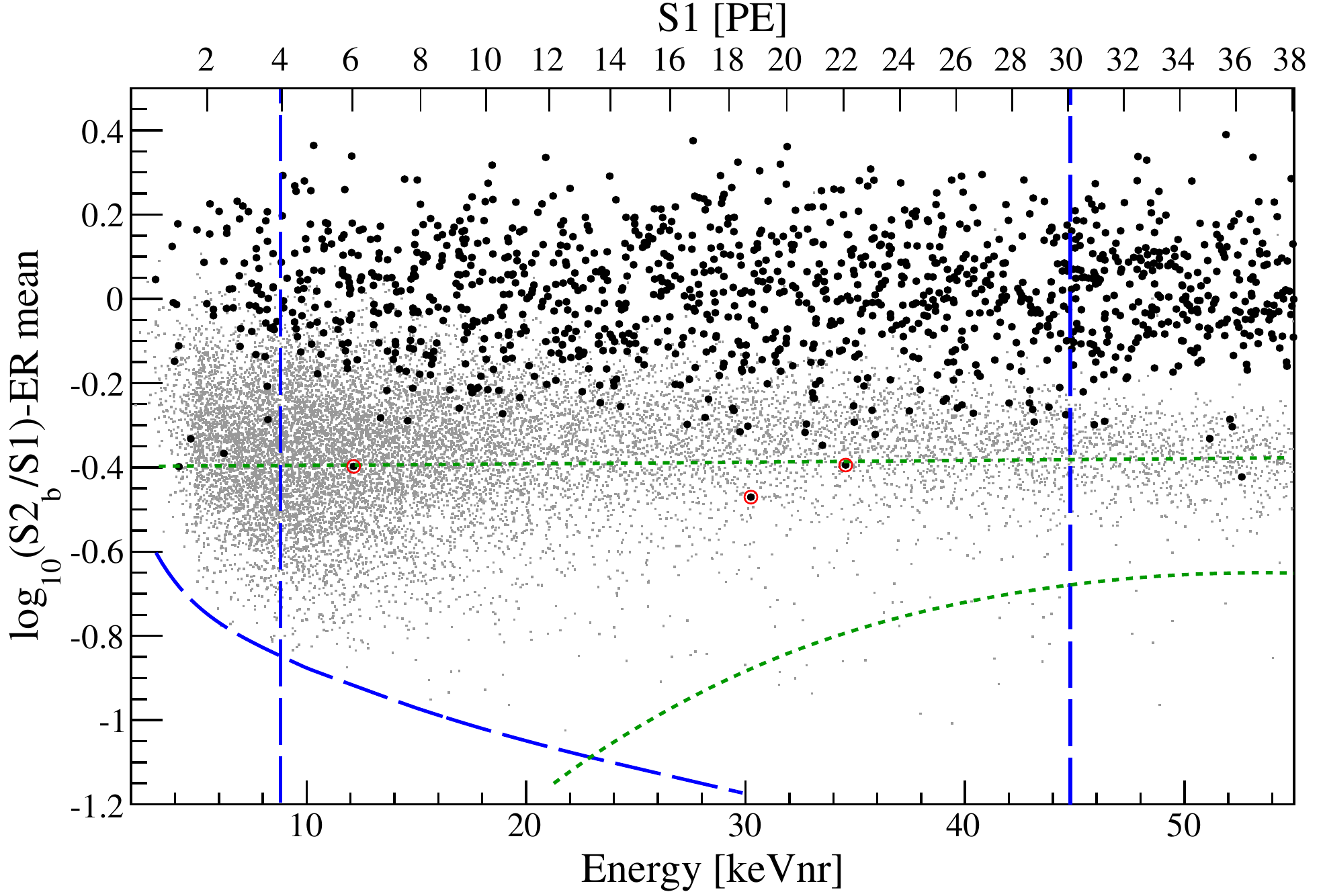}
\includegraphics[width=0.48\columnwidth]{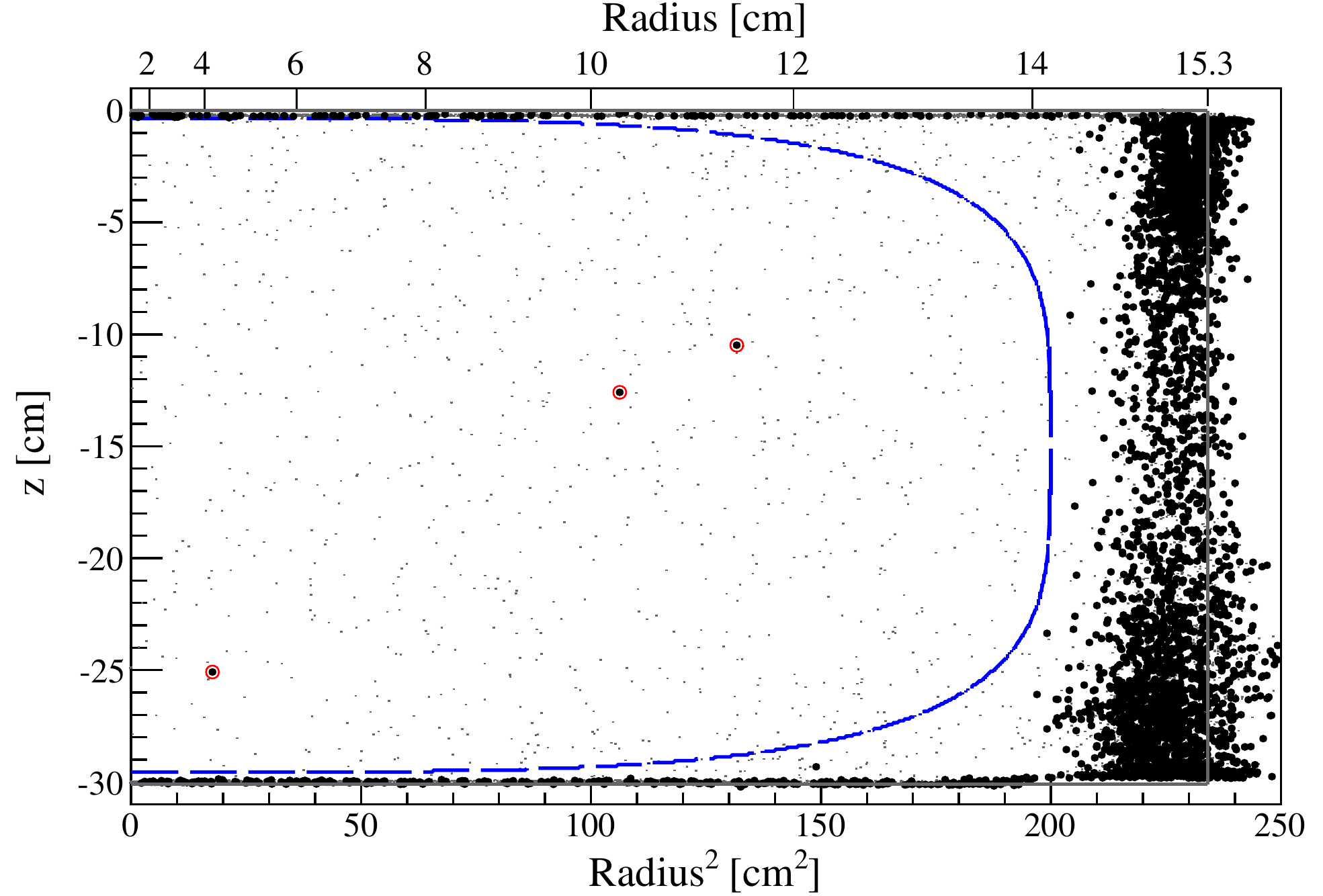}
\caption{{\small (Left): Event distribution in  $\log_{10}(\n{S2/S1})$ as a function of energy.  Gray dots show nuclear recoils as measured with an AmBe source, black dots are the electronic recoil background. The WIMP-search region is defined by the energy window $(8.4-44.6)\1{keV_{nr}}$ and the lower bound of the software threshold $\n{S2}>300\1{photoelectrons}$ (dashed). The optimum interval analysis additionally considers the 99.75\% rejection line from above and the $3\sigma$-contour of the nuclear recoil event distribution from below  (dotted). Three events fall into the WIMP search region (circles), with $(1.8 \pm 0.6)$ events expected from background. (Right): Spatial distribution of all events (gray dots) and events below the 99.75\% rejection line (black dots) in the TPC observed in the $(8.4-44.6)\1{keV_{nr}}$ energy range during 100.9~live days.  The 48~kg fiducial volume (dashed) and the TPC dimensions (gray) are indicated as well \cite {xenon100_prl107}.}}
\label{fig:scatter}
\end{figure}

The most recent {\small XENON100} results \cite{xenon100_prl107} are derived from 100.9~live days of dark matter data acquired between January and June 2010.  A blind analysis, using the so-called {\sl hidden signal box technique}, where events in and around the signal region are kept hidden until the analysis is complete, was conducted. The event selection cuts and the background predictions were fixed before the box was opened, the signal box being defined in a two-parameter space, namely $\log_{10}$(S2/S1) versus energy, as shown in figure \ref{fig:scatter}, left.

\begin{figure}[htbp]
\centering
\includegraphics[width=0.7\columnwidth]{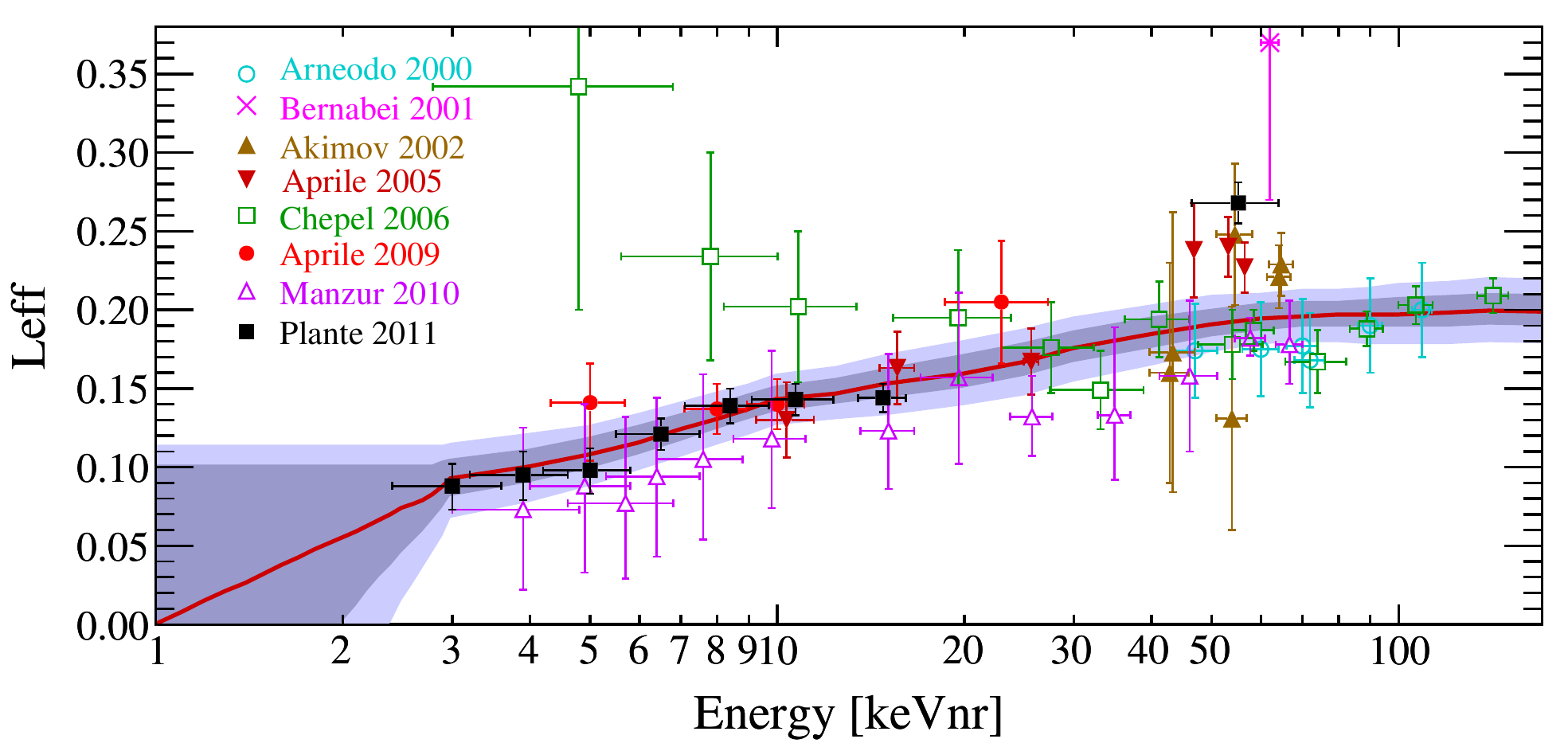}
\caption{{\small Direct measurements of $\mathcal{L}_{\rm{eff}}$ versus recoil energy (\cite{plante11} and references therein). 
The data can be described by a Gaussian distribution, its mean (solid) and the $1\sigma$ and $2\sigma$ uncertainty bands are shown. Below $3\1{keV_{nr}}$
the behaviour is logarithmically extrapolated to zero at $1\1{keV_{nr}}$ \cite {xenon100_prl107}.}}
\label{fig:leff}
\end{figure}

The energy of nuclear recoils is inferred from the S1 signal, as $E_{\n{nr}} =  (S1/L_y) (1/\mathcal{L}_{\rm{eff}})(S_{\n{ee}}/S_{\n{nr}})$. The scintillation efficiency $\mathcal{L}_{\rm{eff}}$ of nuclear recoils relative to that of 122\,keV $\gamma$-rays at zero drift field, used as a ``standard candle",  is  parametrized as shown in figure~\ref{fig:leff}.  It includes recent measurements down to 3\,keV$_{\n{nr}}$ nuclear recoil energy \cite{plante11}, in addition to all other direct measurements.
$\mathcal{L}_{\rm{eff}}$ is logarithmically extrapolated below the lowest measured point, motivated by the trend in the data as well as phenomenological 
studies which simultaneously take into account light and charge signal~\cite{Bezrukov:2010qa}. The electric field scintillation quenching factors are $S_{\n{ee}}=0.58$ and $S_{\n{nr}}=0.95$  for electronic and nuclear recoils \cite{Aprile:2006kx}, 
and the detector's light yield at 122\,keV and  drift field of $530\1{V/cm}$ is  $L_y =(2.20\pm0.09)\1{photoelectrons/keV}$.

The dark matter result was  based on a profile likelihood analysis as introduced for the commissioning run  in~\cite{Aprile:2011hx}, taking into account the relevant backgrounds for this new dataset. The profile likelihood analysis does not employ an event selection cut based on the S2/S1-ratio; moreover,  the signal and background hypothesis are tested {\sl a priori}, regardless of the observed data. As a cross check,  an analysis based on the optimum interval method~\cite{Yellin:2002xd} was performed in parallel. The restricted S2/S1-space  defines a {\small WIMP}-search region which allows a direct comparison of the observed signal with the number of expected background events. 

By comparing the measured background rate in this run with Monte Carlo simulations of the expected electronic recoil background components~\cite{aprile2011vb}, a $^{nat}$Kr/Xe concentration of (700$\pm$100)$\times$10$^{-12}$\,mol/mol,  
was inferred, which is higher than the concentration observed in the commissioning run (see section \ref{sec:backgrounds}). The additional Kr was introduced by an air leak during maintenance work on the gas recirculation pump, prior to the start of the data-taking period. After the science run, the krypton concentration has been reduced by cryogenic distillation to the previous level, as confirmed with a $\beta$-$\gamma$-coincidence analysis.

\begin{figure}[htbp]
\centering
\includegraphics[width=0.75\columnwidth]{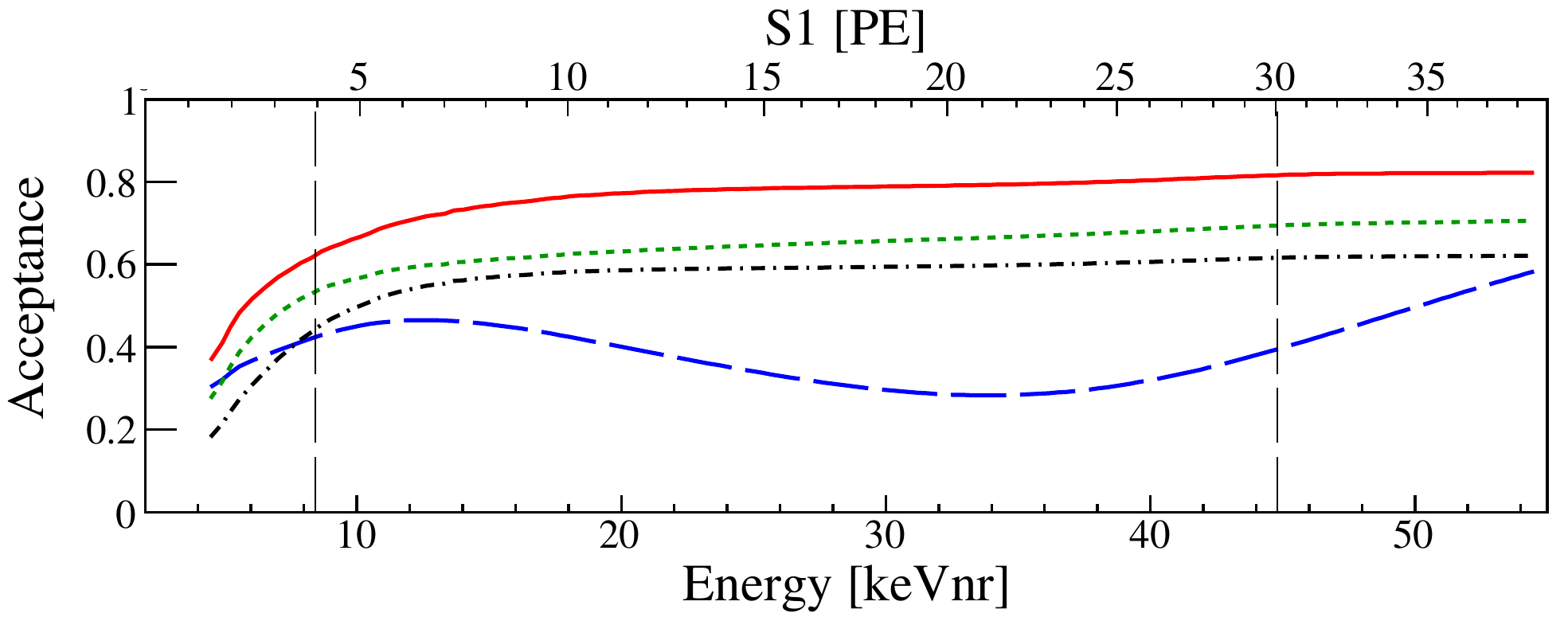}
\caption{{\small Acceptance of all data quality cuts  for $m_W \ge 50$~GeV/$c^2$ (solid), $m_W = 10$~GeV/$c^2$ (dotted), $m_W = 7$~GeV/$c^2$ (dash-dotted). The optimum interval analysis uses an additional event selection cut based on the 
S2/S1-ratio, its acceptance for nuclear recoils is shown as well (dashed) \cite {xenon100_prl107}.}}
\label{fig:acceptance}
\end{figure}

Requirements to the quality and topology of events are designed to retain a high acceptance of the expected WIMP-induced single-scatter nuclear recoils. The majority of selection cuts were designed and fixed before unblinding the signal region, based on expected signal characteristics, on nuclear recoil data from the AmBe calibration, and on low-energy electronic recoils from Compton-scattered gammas. To satisfy the requirement of a WIMP signature to be a localized interaction, one S2~signal above 300 photoelectrons is required, corresponding to about 15~ionization electrons.  The corresponding S1~signal must be above 4 photoelectrons and satisfy a two-fold {\small PMT} coincidence in a $\pm$20~ns window, without having a coincident signal in the LXe veto. Any other S1-like signal must be consistent with electronic noise or unrelated to the S2, based on its S2/S1-ratio. In addition, both S1 and S2 {\small PMT} hit patterns as well as the width of the S2 pulse are required to be consistent with a single interaction vertex at the reconstructed position. The cumulative cut acceptance, used by both analyses, is shown in figure~\ref{fig:acceptance} and has an error of~$\sim3\%$. It is estimated based on Monte Carlo simulations, AmBe and $^{60}$Co calibration data, as well as electronic recoils observed outside the {\small WIMP} search region (in so-called sidebands) during the dark matter run. It includes a {\small WIMP} mass dependent S2-acceptance which is derived from the expected recoil spectrum and the measured S2 versus~S1 distribution.

The  $(8.4-44.6)\1{keV_{nr}}$ energy window for the {\small WIMP}-search, corresponding to $(4-30)$ photoelectrons,  was chosen such as to yield sufficient discrimination between genuine S1~signals and electronic noise at its lower bound,  while including most of the expected {\small WIMP} signal at its upper bound.
Given the homogeneously distributed $^{85}$Kr background, the fiducial volume was optimized on electronic recoil background data  to 48\,kg.
The background rejection level was set to 99.75\% and its acceptance to nuclear recoils calculated using single-scatter nuclear recoils from AmBe data (shown in figure~\ref{fig:acceptance}). The profile likelihood analysis tests the full 
S2/S1-space without using this additional event selection cut.

The expected background in the {\small WIMP}-search region is the sum of Gaussian leakage from electronic recoil background, of non-Gaussian leakage, and of nuclear recoils from neutron interactions. The latter, estimated by Monte Carlo simulations, takes into account neutron spectra and total production rates from $(\alpha,n)$ and spontaneous fission reactions in the detector and shield materials, with input from measured radioactivity concentrations~\cite{Aprile:2011ru}. The muon-induced neutrons are modeled as well,  and contribute 70\% to the total nuclear recoil background \cite{alex_phd_thesis}. Considering the measured trigger efficiency and the energy threshold in the active xenon veto, the overall prediction is $(0.31^{+0.22}_{-0.11})$ single-scatter nuclear recoils in the 100.9~days data, before an S2/S1-cut, in the energy region of interest and 48~kg fiducial xenon mass, of which $(0.11^{+0.08}_{-0.04})$ are expected in the signal box. 

The normalized electronic recoil band, obtained by subtracting its mean as inferred from calibration data, is well described by a Gaussian distribution in $\log_{10}(\n{S2/S1})$ space. Gaussian leakage, dominated by the $^{85}$Kr background, is predicted from the number of background events outside the blinded signal region, taking into account the blinding cut efficiency and the background rejection level. It is $(1.14 \pm 0.48)$~events in the {\small WIMP}-search region. Non-Gaussian leakage, due to double-scatter gamma events with one interaction in a charge insensitive region and another in the active target, is also estimated from calibration data,  yielding $(0.56^{+0.21}_{-0.27})$~events. The total background prediction in the  {\small WIMP}-search region for 99.75\%  rejection, 100.9~days of exposure and 48\,kg fiducial mass is $(1.8\pm0.6)$~events. The profile likelihood analysis uses identical data sets and background assumptions to obtain prediction for the Gaussian, non-Gaussian and neutron background for every point in the $\log_{10}(\n{S2/S1})$ parameter space.

Three events, at energies of $12.1~\1{keV_{nr}}$, $30.2~\1{keV_{nr}}$, and  $34.6~\1{keV_{nr}}$  pass all quality criteria for single-scatter nuclear recoils in the signal region; these are shown in  figure~\ref{fig:scatter}, left, in the normalized  $\log_{10}(\n{S2/S1})$ space. The observation does not depend on moderate variations in the definition of data quality cuts. Their spatial distribution in the TPC is shown in figure~\ref{fig:scatter}, right. Given a background expectation of $(1.8\pm0.6)$ events, no dark matter discovery can be claimed, the chance probability of the corresponding Poisson process to result in 3~or more events being 28\%. Consistent with above result, the  profile likelihood analysis does not see a significant signal excess, the $p$-value of the background-only hypothesis being 31\%. 

\begin{figure}[tbph]
\centering
\includegraphics[width=0.7\columnwidth]{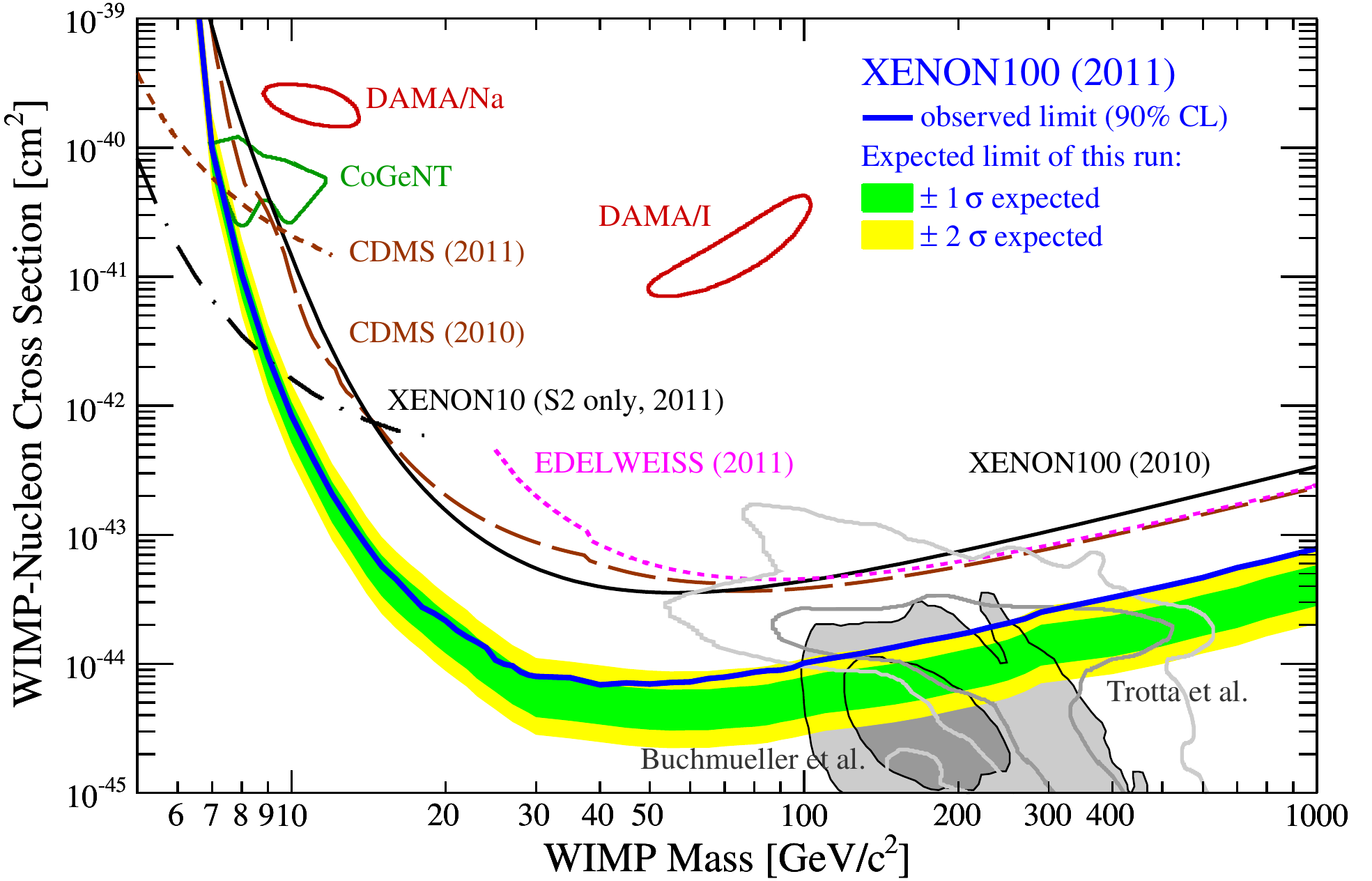}
\caption{{\small XENON100 limit (at 90\%~CL, thick line) on the spin-independent WIMP-nucleon cross-section, together with the expected  sensitivity of the run (shaded bands).  Previous results from XENON100, as well as from other experiments are also shown. Theoretical expectations from supersymmetry are indicated at 68\% and 95\%~CL (shaded gray~\cite{Buchmueller:2011ki} and gray contour~\cite{Trotta:2008bp}) \cite {xenon100_prl107}.}}
\label{fig:limit}
\end{figure}

The experimental upper limit on the scalar  {\small WIMP}-nucleon elastic scattering cross-section, shown in figure~\ref{fig:limit}, is calculated for the standard halo model with $v_0=220$~km/s, an escape velocity of $v_{\n{esc}}=(544^{+64}_{-46})$~km/s, and a local density of $\rho_h=0.3\1{GeV/cm^3}$.  The S1~energy resolution, governed by Poisson fluctuations of the photoelectron generation in the {\small PMTs}, is taken into account. Uncertainties in the energy scale as indicated in figure~\ref{fig:leff}, in the background expectation and in $v_{\n{esc}}$ are profiled out and incorporated into the limit.  A minimum cross section limit of  7$\times10^{-45}\1{cm^2}$, at 90\% C.L., is reached at a WIMP mass of 50$\1{GeV/c^2}$. The impact of $\mathcal{L}_{\rm{eff}}$ below $3\1{keV_{nr}}$ is negligible at a mass of 10$\1{GeV/c^2}$. 
The limit at higher masses is weaker than the expected sensitivity because of  the presence of two events around $30\1{keV_{nr}}$. Within the systematic differences of the methods, the result is consistent with the one from the optimum interval analysis, which calculates the limit based only on events in the {\small WIMP}-search region. Its acceptance-corrected exposure, weighted with the spectrum of a 100$\1{GeV/c^2}$ {\small WIMP}, is $1471\1{kg\times days}$. 
{\small XENON100} thus improves upon {\small XENON10} results, and  starts to probe the region where supersymmetric dark matter is accessible to the LHC~\cite{Buchmueller:2011ki}.

The {\small XENON100} detector is taking new dark matter data with a reduced krypton background, a lower trigger threshold and an improved performance since March 2011. More than 210 live days are currently on disk, results are expected to be released in spring 2011. In parallel, the preparations for {\small XENON1T} are ongoing, and its construction at {\small LNGS} will start in late 2012.

\section{Acknowledgements}
This work has been supported by the National Science Foundation Grants No.~PHY-03-02646 and PHY-04-00596, the Department of Energy under Contract No.~DE-FG02-91ER40688, the CAREER Grant No.~PHY-0542066, the Swiss National Foundation SNF Grant No.~20-118119, the Volkswagen Foundation, the University of Zurich and the FCT Grant No.~PTDC/FIS/100474/2008.

\end{document}